\documentclass[doublespacing]{elsart}

\usepackage{amssymb,graphicx}

\begin{document}
\journal{Physica B}

\begin{frontmatter}
\null
\kern -20 mm
\rightline{\Large ThM4.2I}
\title{Photoluminescence of single colour defects \\ in 50~nm diamond nanocrystals}

\author{F. Treussart\corauthref{cor}\thanksref{ENS}},
\ead{francois.treussart@physique.ens-cachan.fr}
\author{V. Jacques\thanksref{ENS}},
\author{E. Wu\thanksref{ENS}},
\author{T. Gacoin\thanksref{PMC}},
\author{P. Grangier\thanksref{LCFIO}.}, and 
\author{J.-F. Roch\thanksref{ENS}} 
\corauth[cor]{Corresponding author
\quad F. Treussart\ \\
\quad LPQM, \'Ecole Normale SupŽrieure de Cachan\\
\quad61 avenue du pr\'esident Wilson, 94235 Cachan cedex, France\\
\quad tel +33-1-4740-7555; \ fax +33-1-4740-2465}

\address[ENS]{Laboratoire de Photonique Quantique et
Mol\'eculaire, UMR 8537 du CNRS, ENS Cachan, 61 avenue du Pr\'esident Wilson, 94235 Cachan
cedex, France}
\address[PMC]{Laboratoire de Physique de la Mati\`ere Condens\'ee, UMR 7643 du CNRS,
\'Ecole Polytechnique, 91128 Palaiseau cedex, France}
\address[LCFIO]{Laboratoire Charles Fabry de l'Institut d'Optique, UMR 8501 du CNRS, BP 147, 91403 Orsay cedex, France} 

\kern -2mm
\begin{abstract}
We used optical confocal microscopy to study optical properties of diamond 50~nm nanocrystals first irradiated with an electron beam, then dispersed as a colloidal solution and finally deposited on a silica slide. At room temperature, under CW laser excitation at a wavelength of 514.5~nm we observed perfectly photostable single Nitrogen-Vacancy (NV) colour defects embedded in the nanocrystals. From the zero-phonon line around 575~nm in the spectrum of emitted light, we infer a neutral NV$^0$ type of defect. Such nanoparticle with intrinsic fluorescence are highly promising for applications in biology where long-term emitting fluorescent bio-compatible nanoprobes are still missing.
\end{abstract}

\end{frontmatter}

\pagebreak

\section{Introduction}\label{}
A growing interest for single colour defects in diamond~\cite{Gruber_97} has recently emerged for their applications in quantum information processing~\cite{Wrachtrup_PRL04_1,Wrachtrup_PRL04_2} and quantum communications~\cite{Kurtsiefer00,Brouri00}.
In addition to the same high quantum efficiency than dye molecules they have a quasi-perfect photostability at room temperature that makes them better suited for long-term applications.
Contrary to semiconductor quantum dots (QDs)~\cite{Brokmann_PRL03}, photoluminescence from diamond colour centres is very stable and not affected by blinking processes.

Triggered single-photon source based on the temporal control of the photoluminescence of such a single colour centre in a diamond nanocrystal~\cite{Rosa_PRA00} has been already built~\cite{Alexios_EPJD} and then implemented in quantum cryptography testbeds~\cite{Beveratos02,Romain_QKD_NJP04}. Among the single photon sources based on single dipoles fluorescence (atoms~\cite{Brattke_PRL01,Kuhn_PRL02,AntoineB_Science05,Kimble_Science04}, molecules~\cite{Brunel_PRL99,Lounis00,FMT_PRL02} and semiconductor quantum dots~\cite{Pelton_PRL02,Santori_Nature02,Fattal_PRL04_2}) which were developed in the past few years for application to quantum information processing, the diamond colour centre source has the advantage of simplicity, long-term stability and room temperature operation.
 
Up to now, the smallest diamond nanocrystals used had a mean size of about 90~nm~\cite{Beveratos01}. 
In this article, we show that photostable single colour centre with a maximum photoluminescence intensity centered at 630~nm can be observed in 50~nm nanocrystals obtained by finer grinding of the original micron size diamond powder. 
Taking into account the possibility of their surface functionalization~\cite{Krueger_Carbon05}, this makes such photoluminescent nanocrystals probably a good alternative to other existing nanoparticles (e.g. QDs) for biological applications like targeted drug delivery inside living cells. 

\section{Sample preparation}
We start from commercial type Ib synthetic monocrystaline diamond powder (\emph{Syndia SYP 0--0.05}, from Van Moppes \& Sons, Geneva) fabricated for industrial polishing purposes. This powder is produced from raw monocrystalline material by crushing, purification, pregrading and a final precision grading to achieve a particle size distribution below 50~nm (99\% confidence interval).

Colour centres are associated to atom impurities in the diamond matrix. The best known defect is the Nitrogen-Vacancy (NV) defect corresponding to the association of a substitional nitrogen atom with  a vacancy in an adjacent crystalline site. The synthetic diamond we started from is supposed to have a sufficiently high nitrogen contents, up to 300 ppm~\cite{Zaitsev_book}. It is then necessary to create the associated vacancies. We therefore irradiate the powder with 1.8~MeV electrons at a dose of 10$^{17}$~electrons/cm$^2$. Irradiation is then followed by annealing at 800~$^\circ$C during 2 hours, to induce stable NV centre formation~\cite{Gruber_97,Brouri00}.

\begin{figure}
\includegraphics[width=\textwidth]{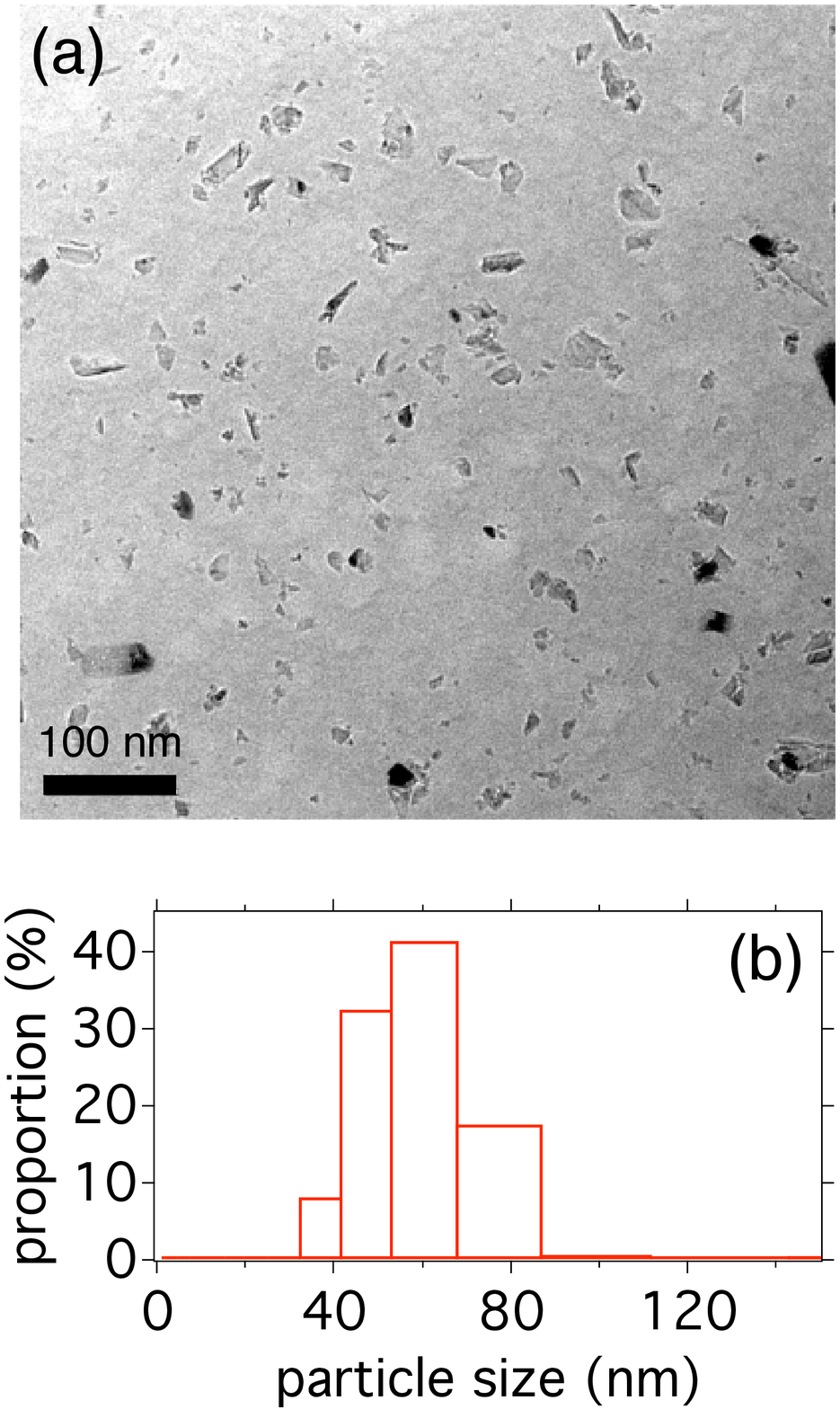}
\caption{}
\label{fig:fig1}
\end{figure}

The irradiated powder is then dispersed in a 1~\% in weight polyvinylpyrrolidone solution in isopropanol to achieve a stable colloidal solution of concentration 2~g/l. Ultracentrifugation at a speed of 11000 rpm for 40 min allows us to further select the smallest particles. Size measurement by dynamic light scattering technique (DLS) gives a size estimate of  50~nm (see fig.\ref{fig:fig1}(b)). 
This value is consistent with the observation of the particles by transmission electron microscopy 
as shown is figure~\ref{fig:fig1}(a). The particles appear to be well dispersed, with an angular shard like shape, as expected considering the crushing process used for the elaboration of the initial crude diamond powder. 
  
\section{Photoluminescence of single NV$^0$ colour centre}
To study of the photoluminescence properties of the diamond nanocrystals photoluminescence we spincoated the solution on silica (Suprasyl) polished slides with thickness $\approx 0.15$~mm. Fluorescence of the sample was then investigated with a home-made inverted confocal microscope using a setup described in more details in Ref.\cite{Treussart_JLumin04}. A CW argon laser beam at a wavelength of 514.5~nm was used to excite the NV colour centre. The laser beam was focused onto the sample with an oil immersion microscope objective ($\times 100$, numerical aperture 1.4) which also collects the emitted photons. Optical filters separate the excitation laser light from the photoluminescence centered at a wavelength of about 630~nm. The later is detected with silicon avalanche photodiodes in single photon counting regime. 
Part of the photoluminescence light is sent towards an imaging spectrograph equipped with a cold CCD back-illuminated matrix.

\begin{figure}
\includegraphics[width= \textwidth]{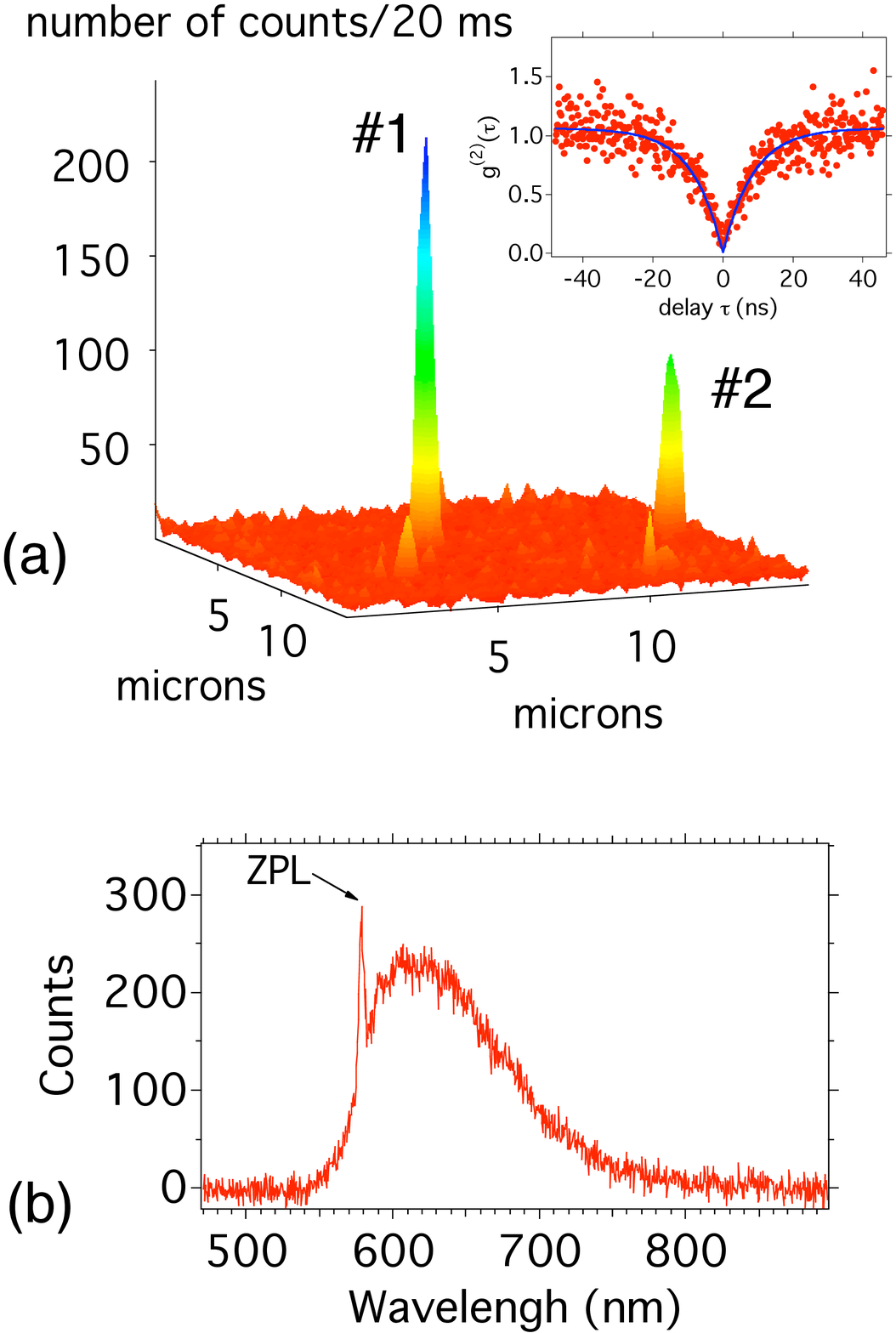}
\break
\caption{}
\label{fig:fig2}
\end{figure}

To identify well isolated photoluminescent emitters, we first raster scan the sample using CW excitation (figure~\ref{fig:fig2}(a)). From the most intense peak (\#1), we deduce a signal over background ratio of about 30. 
For each photoluminescent spot, the unicity of the emitter is then checked by the observation of antibunching in the photoluminescence intensity~\cite{Brunel_PRL99}. Since after the emission of a first photon, it takes a finite time for a single emitter to be excited again and then spontaneously emits a second photon, the antibunching effect appears as a dip around zero delay $\tau=0$ in the normalized intensity autocorrelation function~\cite{Brouri00}
\begin{equation}
g^{(2)}(\tau) \, \equiv \, \frac{\langle I\left(t\right)I\left(t+\tau\right)\rangle}{{\langle
I\left(t\right)\rangle}^2}.
\label{autocorrelation}
\end{equation} 
The later function is deduced from the histogram of time delays between two consecutively detected photons. Due to the existence of a 30~ns photodetection deadtime on the avalanche photodiode, two photodiodes placed on both ports of a non-polarizing beamsplitter cube are required for this measurement following the standard Hanbury Brown and Twiss scheme~\cite{Mandel_Wolf}. Considering our detection efficiency of a few percent, this histogram is a very good approximation of the intensity autocorrelation function $g^{(2)}$~\cite{Reynaud_these_etat}. 

The inset of Figure~\ref{fig:fig2}(a) shows the normalized $g^{(2)}$ function corresponding to centre \#1. The dip at zero delay is a clear evidence of photon antibunching in the emitted light and the signature of a single colour centre. The very small remaining value at zero delay $g^{(2)}(0)\approx 0.06$ is due to background photoluminescence light from the substrate and from the diamond nanoparticle embedding the centre itself.

Centre \#1 spectrum displayed on figure~\ref{fig:fig2}(b) exhibits a zero-phonon line at 579~nm which is attributed to neutral NV centre~\cite{Mita96,Iakoubovskii00,Treussart_JLumin04}. 
The creation of such centre during the irradiation process, instead of the most dominant negatively charge form NV$^-$ is probably due to a too high electron dose as suggested in Ref.\cite{Mita96}. When the irradiation dose is high enough so that all the nitrogen have been converted either in N$^+$ or in NV$^-$ (having its zero-phonon line at 637~nm), creation of additional vacancy by further irradiation results in the conversion of NV$^-$ into NV$^0$. Note that such a transformation was also observed under short pulse laser illumination of single diamond nanocrystals containing initially NV$^-$ defects~\cite{Treussart_JLumin04}.

\begin{figure}
\includegraphics[width= \textwidth]{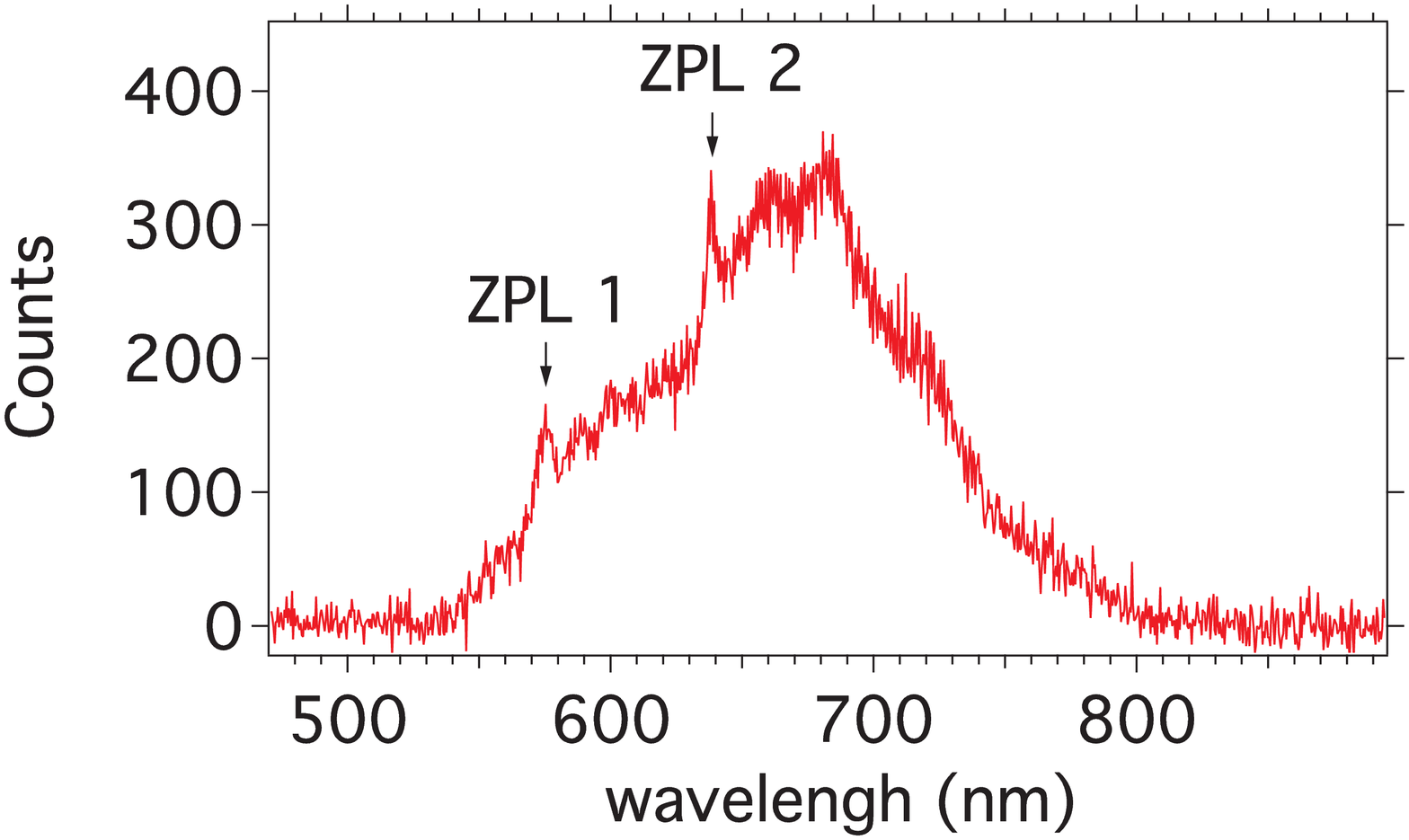}
\caption{}
\label{fig:fig3}
\end{figure}

Peak \#2 displayed in figure~\ref{fig:fig2}(a) raster scan also corresponds to a single NV$^0$ centre. Summarizing, among the 22 studied emitting spots, 9 were single NV$^0$ colour centres. The others had a flat $g^{(2)}$ function normalized to one (no antibunching) meaning that they were not single. For most of the later emitters, the spectra extend at longer wavelength with both zero-phonon lines of neutral and negatively charged NV centres as shown on figure~\ref{fig:fig3}. The presence of NV$^-$ features is tentatively related to the largest host nanocrystals present in the sample for which the number of embedded nitrogen impurities is high enough for not having the complete conversion of NV$^-$ into NV$^0$ centres under irradition.

\section{Conclusion}
We have observed stable photoluminescence of single colour centre in 50~nm diamond nanocrystals from commercial abrasive powder. Using appropriate chemical processing, the diamond nanoparticles can be dispersed in a colloidal solution. From the spectrum of the emitted light, those centres appear to be of the Nitrogen-Vacancy neutral type. We believe that the high dose of irradiation used transform the NV$^-$ centres first created into NV$^0$. Complementary experiments are now undertaken to check this interpretation, by varying the irradiation dose.

Further decrease in size of the nanocrystal by additional untracentrifugation will be done in order to achieve a particle size of about 10~nm, in the range of other fluorescent probes used for biological applications, like CdSe/ZnS QDs. Photoluminescent diamond nanocrystals could therefore become, in a near future, a promising alternative to existing fluorescent nanoprobles and drug delivery device.

\section*{Acknowledgements}
This work was partly supported by a ``AC Nanosciences'' grant from Minist\`ere de la Recherche, and Institut Universitaire de France. The authors would like to thank S.~Esnouf (Laboratoire des Solides irradi\'es, \'Ecole Polytechnique) for the electron irradiation.

\newpage
\section{Figure Captions}

\paragraph*{Figure 1} (a)TEM microscope image of diamond nanocrystals colloidal solution spincoated on an electronic microscope grid. Angular shape of most of the nanocrystals is due to crushing in the fabrication process leading to cutting edge suited for abrasive applications of this diamond powder.
(b) Histogram of colloidal nanoparticles size distribution in the polymer solution, obtained from dynamic light scattering (DLS) measurement. A mean size of 50~nm is inferred from the later measurement, larger than the estimated size from TEM image. The difference comes from the fact that the DLS measurement takes into account the polymer molecules adsorbed at the surface of the diamond particle.

\paragraph*{Figure 2}
{\bf (a)} Photoluminescence raster scan of diamond nanocrystals spincoated on a silica slide, excited with CW argon laser at a wavelength of 514.5~nm and a power $P=5$~mW. Half of the photoluminescence was sent towards the spectrograph and only one of the two detectors involved in time intensity correlation measurement was used to record the raster scan. An overall counting rate of about 40~kcounts/s can therefore be deduced. We checked that the colour centre photoluminescence was saturated for this excitation power.
\emph{Inset:} Second order intensity autocorrelation function $g^{(2)}(\tau)$ of the emitted photons (excitation power $P=6$~mW). The curve shows a 671~s lasting record of the time delays $\tau$ between two consecutively detected photons on the two avalanche photodiodes. The dip at $\tau=0$ (antibunching) gives evidence for single emitter photoluminescence, while $g^{(2)}=1$ corresponds to the shotnoise limit for a poissonian photon distribution. For time delays longer than a few ten nanoseconds,  $g^{(2)}$ is slightly larger than one. This bunching behavior comes from intermittency in the light emitted, due to shelving in a dark level~\cite{Brouri00}. Fit of the data by a symmetric exponential function (solid line, in blue) yields a characteristic time of 9.7~ns.
{\bf (b)} Spectrum of the \#1 single colour centre shown in the figure (a), for CW excitation at a wavelength of 514.5~nm, and excitation power of 5~mW. Integration duration on the CCD matrix, cooled at -65$^\circ$C is 4 minutes. Note that both the tranmission of the filters placed on the detection path, and the detector quantum efficiency are flat in the wavelength range of interest. ZPL indicates the well defined narrow zero-phonon line at 579~nm, which is related to an NV colour centre of the \emph{neutral} form NV$^0$.

\paragraph*{Figure 3}
Spectrum of light from a photoluminescent spot with \emph{no associated antibunching}, it therefore corresponds to the emission of more than about 5 centres. The spectrum exhibits two ZPL: ZPL~1 (575~nm) reveals the presence of NV$^0$ colour centres, while ZPL~2 (637~nm)  gives evidence for negatively charged NV$^-$ centres. The integration duration is 4 minutes, under CW laser irradiation at 5~mW excitation power.

\end{document}